\documentclass[aps,prb,twocolumn,groupedaddress,showpacs,superscriptaddress,amssymb,amsmath]{revtex4-2}
\usepackage{graphicx}
\usepackage{dcolumn}
\usepackage{bm}
\usepackage{hyperref}
\usepackage{cleveref}
\hypersetup{
    colorlinks=true,
    linkcolor=blue,
    urlcolor=blue,
	citecolor=blue
}

\usepackage{comment}
\usepackage{color}
\usepackage[utf8]{inputenc}
\usepackage{graphicx}
\usepackage{tabularx}
\usepackage{xcolor}
\usepackage{amsmath}
\usepackage{dcolumn}
\usepackage{hyperref}
\usepackage{bm}
\usepackage{epsf}
\usepackage{braket}
\usepackage{tensor}
\usepackage{soul}

\newcommand\la{\langle}
\newcommand\ra{\rangle}

\begin{document}
\newcolumntype{M}[1]{>{\centering\arraybackslash}m{#1}}

\title{Quantum Dynamics of Full Counting Statistics in Fermionic Lattices with Localized Gain}

\author{Bijay Kumar Agarwalla}
\affiliation{Indian Institute of Science Education and Research Pune, Maharashtra, 411008, India}

\author{Manas Kulkarni}
\affiliation{International Centre for Theoretical Sciences, Tata Institute of Fundamental Research,
Bengaluru 560089, India}

\date{\today}

\begin{abstract} 
We investigate the dynamics of full counting statistics (FCS) of the growth of the total number of fermions in a one-dimensional non-interacting lattice subjected to a localized particle gain (source) at one edge. The dynamics of the setup is modeled by the Gorini--Kossakowski--Sudarshan--Lindblad (GKSL)  quantum master equation. We recast the 
counting problem that involves counting particle number at every lattice site to a problem where only the local injected site is involved.
We employ the Schwinger-Keldysh path integral formalism within the GKSL framework and derive an analytical expression for the cumulant generating function at arbitrary times and obtain a Levitov-Lesovik type formula in the long-time limit, earlier derived for boundary driven setups in the steady-state. For clean lattices with either short- or long-range hopping, supporting single-particle delocalized eigenstates, we show that all cumulants grow linearly with time in the asymptotic regime. Our analytical results are in excellent agreement with direct numerical simulations. These findings establish a general framework for characterizing FCS and quantum fluctuations in driven open fermionic systems with localized particle injection.
\end{abstract}

\maketitle

\textit{Introduction.--} 
The full probability distribution of observables in nonequilibrium quantum systems provides detailed information about the underlying microscopic processes and dynamics, extending far beyond that contained in average values~\cite{LL96,LKMP24}. Determining such distributions, however, remains a challenging analytical and numerical problem \cite{danielFCS}. In the context of electron and photon transport, this problem is commonly formulated within the framework of full counting statistics (FCS)~\cite{Bijay2015,Agarwalla_2017}. FCS has emerged as a powerful tool for characterizing quantum fluctuations in mesoscopic transport~\cite{Dario_Landi2022,Chien2015QuantumTI,PhysRevA.85.041601,Klich_2014,Nazarov2003}, where the probability distribution of the integrated particle or charge current has been obtained analytically for several non-interacting steady-state systems. Complementing these analytical developments, numerical techniques such as the density matrix renormalization group~\cite{PhysRevLett.69.2863,PhysRevB.72.180403,dmrg_fcs}, tensor network involving matrix product states~\cite{mps_fcs1,mps_fcs,QGF_prosen,yadalam2026,MZNov2004,itensor,USJan2011,Prosen_QGF1,ganguly2026,moca2026}, have enabled the computation of full probability distributions in interacting many-body systems. Remarkably, under fairly general conditions, the resulting distributions satisfy universal fluctuation relations~\cite{rmp09,Campisi_review,Gallavotti}, which constitute nonequilibrium analogues of the equilibrium detailed balance condition.

A significant focus for FCS problem in open quantum systems has been devoted towards understanding steady-state scenario \cite{Klich2003, KS07,Dhar2007,Levitov1993,sreejith_sandipan} and comparatively much less attention is given to understanding the time dynamics~\cite{dyn_fcs1,dyn_fcs2,dyn_fcs4}, more so from an analytical point-of-view.  In this work, we focus on FCS for particle number growth in the fermionic lattice when it is subjected to a local gain. We model the setup by the Gorrini-Kosakawaski-Sudarshan-Lindblad (GKSL) quantum master equation \cite{breuer2002theory,dutta2025}. In this regard, in a recent set of studies, the impact of a local gain in filling a lattice is studied for both fermions and bosons \cite{Spohn,Krapivsky_2019, Krapivsky_2020,Trivedi2023,6frd-chqr}.
It is also important to note that there has been several studies
on the ``dual" problem, i.e., a system that is subjected to a localized loss~\cite{ultracold_local,PRA_2011,PRL_Zeno,PRA_2013,SELS2020168021, PRA_2017,muller,PRB_2022,doi:10.1126/sciadv.aat6539,loss2024,Bistability_loss,PhysRevLett.129.056802}.

In the case of local gain, it was observed that for tight-binding fermionic lattice, due to the Pauli exclusion principle the growth of fermions scales linearly with time with the rate of growth that depends non-monotonically~\cite{Krapivsky_2019,Trivedi2023} with the injection rate. However, the behaviour of higher order fluctuations has not been investigated. Here, we bridge a gap by studying the quantum dynamics of FCS for the particle number growth in a generic non-interacting one-dimensional lattice.  Starting with an empty lattice subjected to a localized gain term, we ask (i) what is the cumulant generating function corresponding to the FCS of particle number? and 
(ii) how do the higher order cumulants of particle number grow in time asymptotically?

\vspace{0.2cm}
\textit{Setup and quantity of interest.--}
We begin with an arbitrary non-interacting one-dimensional fermionic lattice which is initially empty and is coupled at one edge with a local injection source that injects fermions with rate $\Gamma$. The dynamics of the setup is modelled using the GKSL quantum master equation \cite{Krapivsky_2019,dutta2025}
\begin{equation}
\frac{d \rho}{dt}=-i \big[H, \rho \big] + \Gamma \Big[c_1^{\dagger} \rho c_1 - \frac{1}{2} \big\{c_1 c_1^{\dagger}, \rho\big\}\Big],
\label{GKSL-injection}
\end{equation}
where $H= \sum_{i,j=1}^{L} h_{ij} c_i ^{\dagger} c_j $ is a generic quadratic Hamiltonian of a one-dimensional lattice of size $L$. $h_{ij}$ is the $L \times L$ single particle Hamiltonian matrix. $c_i$ ($c^{\dagger}_i$) is the annihilation (creation) operator for fermions at site $i$.  We are interested in computing the time dynamics of the FCS corresponding to the total fermion number growth in the initially empty lattice. The total fermion number operator is given by $N(t) = \sum_{i=1}^{L} c_i^{\dagger}(t) c_i(t)$.  A convenient way to know about the  FCS is to obtain either the moment generating function (MGF) or the cumulant generating function (CGF), which we focus below.

\vspace{0.2cm}
\textit{Moment Generating function for number statistics.--} We construct the MGF as ${\cal Z}(\lambda,t) = \langle e^{i \lambda N(t)} \rangle$, where the counting parameter $\lambda$ keeps track of the particle number growth in the lattice. The symbol $\langle \cdots \rangle$ represents the average which is taken over  initially empty state of the lattice. As the setup in Eq.~\eqref{GKSL-injection} is quadratic \cite{Prosen_2008}, the MGF can be expressed as a determinant involving the two-point correlation matrix $C_{jk}(t)=\langle c_j^{\dagger} (t)  c_k (t) \rangle$ as \cite{PhysRevLett.134.067101,Klich2003} 
\begin{equation}
{\cal Z}(\lambda,t)= \mathrm{det}\Big[\delta_{jk} + \big(e^{i\lambda}-1\big) C_{jk}(t) \Big]_{j,k=1}^{L}.
\label{GF-corr}
\end{equation}
The equation of motion for the correlation matrix $C_{jk}(t)$ can be obtained following the GKSL equation in Eq.~\eqref{GKSL-injection}. For the given setup it is given by the Lyapunov equation \cite{Prosen_2008,archak_lyap}
\begin{equation}
\frac{dC}{dt}= i \, h_e \, C - i \, C \,  h^{\dagger}_e + G,
\label{Lyapunov}
\end{equation}
where $C$ is the $L \times L$ correlation matrix, $h_e= h + i \,\tilde{\Gamma}/2$ is an effective non-Hermitian matrix, where $\tilde{\Gamma}_{ij}= \Gamma \, \delta_{i1}\, \delta_{j1}$ and $G_{ij} = \Gamma \delta_{i1} \delta_{j1}$, are the  damping and the gain matrices, respectively whose only  (1,1) element is non-zero.  With the help of Eq.~\eqref{Lyapunov}, the MGF in Eq.~\eqref{GF-corr} can be numerically computed and all order moments of $N(t)$  can be easily obtained.  Note that one can alternatively write the CGF from the MGF as,
\begin{eqnarray}
\chi(\lambda,t)&\equiv&\ln {\cal Z}(\lambda,t) \!\!=\!\ln \mathrm{det}\Big[\delta_{jk} + \big(e^{i\lambda}-1\big) C_{jk}(t) \Big]_{j,k=1}^{L}, \nonumber \\
&=& {\rm Tr} \ln \Big[\delta_{jk} + \big(e^{i\lambda}-1\big)  C_{jk}(t) \Big]_{j,k=1}^{L}. 
\label{CGF}
\end{eqnarray}
Eq.~\eqref{CGF} directly yields all cumulants via~\cite{KS07}
\begin{equation}
\kappa_n(t)=
\left.
\frac{\partial^n \chi(\lambda,t)}
{\partial (i\lambda)^n}
\right|_{\lambda=0},
\qquad n=1,2,\ldots,
\label{eq:kappa}
\end{equation}
with $\kappa_1$ and $\kappa_2$ corresponding to the mean and variance, respectively. Note that  in this approach to obtain an explicit analytical expression for the MGF or the CGF, all the components of the $L \times L$ correlation matrix $C(t)$ needs to be evaluated as a function of $t$, which is a non-trivial task. In what follows, we provide an alternate approach and express the CGF  in terms of a local correlation function corresponding to the injected site. This further helps us to obtain an explicit analytical expression of the CGF in the long-time limit. 

\vspace{0.2em}
\textit{Recasting the MGF in terms of properties of the injected site.--} 
We now express the MGF ${\cal Z}(\lambda,t) = \langle e^{i \lambda N(t)}\rangle$ in a form that is amenable for analytical calculations. Since $N(t) = \sum_{i=1}^{L} n_i(t)$ is a sum of local density operator, therefore the counting field $\lambda$ keeps track of   local density at every site of the lattice at each instant of time. However, as change in particle number in the lattice happens only through a localized region, it remarkably turns out that such a global counting problem can be recasted to a counting problem where $\lambda$ keeps track of local particle injection process at the injected site. We can write [see Supplemental material \cite{supp} for the details] 
\begin{equation}
{\cal Z}(\lambda,t) =  \langle e^{i \lambda N(t)}\rangle  = {\rm Tr}\big[e^{L_{\lambda} (t-t_0)} \, \rho (t_0)\big],
\label{alternate-GF}
\end{equation}
where $\rho(t_0)$ is the initial density matrix, and $L_{\lambda}$ is a dressed Liouvillian and is dressed by the counting field $\lambda$. Following the GKSL equation in Eq.~\eqref{GKSL-injection}, the dressed Liouvillian can be expressed as 
\begin{equation}
L_{\lambda} * =-i \big[H, * \big] + \Gamma \Big[e^{i \lambda} \, c_1^{\dagger} * c_1 - \frac{1}{2} \big\{c_1 c_1^{\dagger}, *\big \}\Big],
\label{GKSL-injection-shifted}
\end{equation}
where recall that the first site is coupled to the source. The counting field dresses only the first term in the dissipator, which is often known as the jump term and is responsible for increasing electron number in the lattice.  In what follows, we employ the Schwinger-Keldysh path-integral technique to express the MGF in terms of a Gaussian action which can be integrated exactly and allows to obtain a compact analytical expression in the long-time limit.

\begin{figure*}[t]
\includegraphics[width=\linewidth]{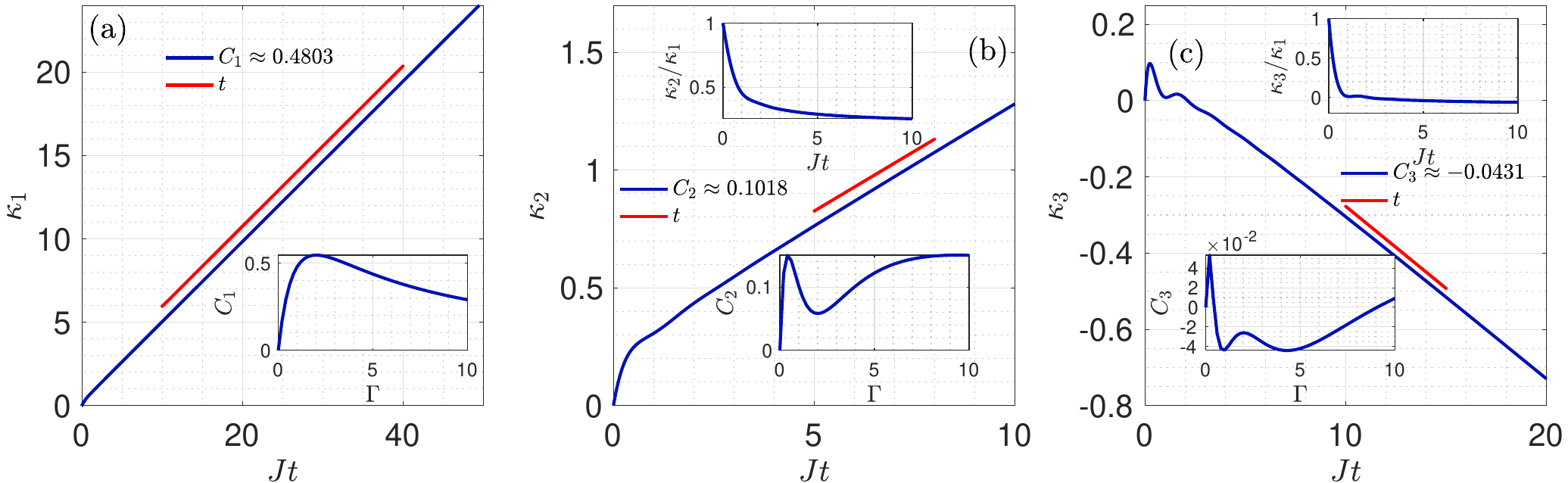}
    \caption{Time dependence of the first three cumulants: (a) $\kappa_1$, (b) $\kappa_2$, and (c) $\kappa_3$ for the non-interacting tight binding lattice described by the Hamiltonian in Eq.~\eqref{eq:ham} with $\alpha \to \infty$. The cumulants are obtained from the cumulant generating function in Eq.~\eqref{CGF} and using Eq.~\eqref{eq:kappa}.  At short times, all cumulants increase linearly with a common rate $\Gamma$, as clearly seen in the insets (upper) of (b) and (c). In the long-time limit, the cumulants continue to grow linearly, indicating constant asymptotic growth rates $\kappa_n= C_n t$, where $n=1,2,3$ indicate different cumulants. Both the short and the late time slopes for all the cumulants match perfectly with analytical predictions as given in Eq.~\eqref{short}, and Eq.~\eqref{CGF-long-time}, respectively. The lower insets in (a)-(c) show the dependence of asymptotic growth rate $C_n$ for $n=1,2,3$ as a function of injection strength $\Gamma$. The plots show a non-monotonic $\Gamma$ dependence.  Parameters used for simulating Eq.~\eqref{CGF} are $L=300$, $\Gamma=1$, for the main plot.}
\label{fig:tight-binding}
\end{figure*}

\vspace{0.2cm}
\textit{Full counting statistics following Schwinger-Keldysh path integral formalism.--} We express $\mathcal{Z}(\lambda,t)$ defined in Eq.~\eqref{alternate-GF} following the path integral formalism as~\cite{kamenev2011field,rammer2007quantum,daniel2025,Aashish2023_KFT}
\begin{eqnarray}
    \mathcal{Z}(\lambda,t) &=& \int \! \! \! D\big[\psi^{*}_{+},\psi_{+},\psi^{*}_{-},\psi_{-}\big]\,e^{i S(\lambda,t)}\, \times \nonumber \\
    &&  \qquad \big \langle \psi_{+}(t_0)|\rho(t_0)|\psi_{-}(t_0)\big \rangle,
     \label{partition_function-1}
\end{eqnarray}
where $|\psi_{\pm} \rangle$ is a column vector at a particular time instant and it contains $|\psi_i\rangle, i=1,2, \cdots L$ with $|\psi_i\rangle$ being the eigenvector (coherent state or Grassmann number for fermions) for the annihilation operator $c_{i}$ with corresponding eigenvalue $\psi_i$ \cite{kamenev2011field}.
$S(\lambda,t)$ is the counting field dressed action given as  (see Supplemental Material.~\cite{supp} for the details of the derivation),
\begin{align}
S(\lambda,t)= \int_{t_0}^{t} d\tau \, \Big[\psi^{*}_{+}& \,\big(i \, \partial_{\tau} \big) \, \psi_{+} - \psi^{*}_{-}\,  \big(i \, \partial_{\tau} \big)\psi_{-}\, \nonumber\\& - i\,f_{\lambda}(\psi^{*}_{+},\psi_{+},\psi^{*}_{-},\psi_{-})\Big], \label{action1}
\end{align}
where the term $f_{\lambda}(\psi^{*}_{+},\psi_{+},\psi^{*}_{-},\psi_{-})$ contains sum of two contributions  $f_{\lambda} =f_u+f^{\lambda}_d$,  one originating from the unitary part $f_u$ and the other from the dissipative part $f^{\lambda}_d$. We have suppressed the arguments for notational simplicity.  The contribution due to the unitary part is given by,
\begin{align}
f_u \!=\!-i\!\!\sum_{i,j=1}^{L}h_{ij}\Big[\psi^{i*}_{+}\psi_{+}^{j}\!-\!\psi_{-}^{i*}\psi_{-}^{j}\Big]. 
\label{fH}
\end{align}
The dissipative contribution due to the local injected source is given by,
\begin{align}
&f^{\lambda}_d= \Gamma \Big[e^{i\lambda}\psi^{1*}_{+}\psi_{-}^{1}+\, \frac{1}{2} \psi^{1*}_{+}\psi_{+}^{1}+\,\frac{1}{2} \psi^{1*}_{-}\psi_{-}^{1}\Big].
\label{dissipative}
\end{align}
Using Eq.~\eqref{fH} and Eq.~\eqref{dissipative}, we can express the action $S(\lambda,t)$ in Eq.~\eqref{action1} as
\begin{align}
\!\!S(\lambda,t)= \int_{t_0}^{t} d\tau \! \int_{t_0}^{t} d\tau' \, \begin{pmatrix}
        \psi^{*}_{+} & \psi^{*}_{-} \\
\end{pmatrix}_{\tau}
\Big[\overline{D}_{\lambda}(\tau,\tau')\Big]
 \begin{pmatrix}
        \psi_{+} \\
        \psi_{-} \\
    \end{pmatrix}_{\tau'}, \label{action}
\end{align}
where  
\begin{equation}
\overline{D}_{\lambda}(\tau,\tau') = \overline{{D}}(\tau,\tau')+{\overline{\Sigma}}_\lambda (\tau,\tau').
\label{Dlambda}
\end{equation}
The different components of the  $\lambda$ independent matrix $\overline{{D}}(\tau,\tau')$ that appears in Eq.~\eqref{Dlambda} are organized as  
\begin{equation}
\overline{D}(\tau,\tau')=
\begin{pmatrix}
    D_{++} & D_{+-}\\
   -D_{-+} & -D_{--}
\end{pmatrix}
\label{D_0matrix}
\end{equation}
with each entry of $\overline{{D}}(\tau,\tau')$ in Eq.~\eqref{D_0matrix}  being of size $L \times L$. These entries are given by
\begin{eqnarray}
\label{D0++}
{D}_{++}(\tau,\tau')&=& \Big[i\, \partial_{\tau} I - h - i \frac{\Gamma}{2} P_1 \Big] \delta (\tau-\tau'), \\
\label{D0+-}
{D}_{+-}(\tau,\tau') &=&  - i \Gamma P_1  \, \delta (\tau-\tau'),  \\
\label{D0-+}
{D}_{-+}(\tau,\tau') &=& 0,  \\
\label{D0--}
{D}_{--}(\tau,\tau') &=& \Big[i \, \partial_{\tau} I - h + i \frac{\Gamma}{2} P_1 \Big] \delta (\tau-\tau').  
\end{eqnarray}
The matrix $P_1$ appearing in Eq.~\eqref{D0++}, Eq.~\eqref{D0+-}, and Eq.~\eqref{D0--} is a $L\times L$ matrix with only $(1,1)$ element being unity and all other elements are zero. The matrix $\overline{D}$ in Eq.~\eqref{D_0matrix} that involves differential operators can be associated with the Green's functions by satisfying the integral operator equation
 \begin{equation}
 \int_{t_0}^{t} d\tau'' \, \overline{D}(\tau, \tau'') \, \overline{G}(\tau'', \tau') = I \, \delta(\tau-\tau')
 \label{integro-diff}
 \end{equation}
The different components of the $\lambda$ dependent matrix ${\overline{\Sigma}}_\lambda$ in Eq.~\eqref{Dlambda}  are given by 
\begin{eqnarray}
\big[\Sigma_{\lambda}(\tau,\tau')\big]_{++} &\!=\!& \big[\Sigma_{\lambda}(\tau,\tau')\big]_{-+} \!=\!\big[\Sigma_{\lambda}(\tau,\tau')\big]_{--} \!=\! 0, \nonumber  \\
\big[\Sigma_{\lambda}(\tau,\tau')\big]_{+-} &=& -i \, \Gamma (e^{i \lambda}-1) \,  P_1 \, \delta(\tau-\tau'). 
\label{Sigma_lambda}
\end{eqnarray}
Using the Gaussian action in Eq.~\eqref{action} and starting with empty lattice as an initial state, the MGF in Eq.~\eqref{partition_function-1} corresponds to a multi-variable Gaussian integration in terms of the Grassmann variables that results in 
\begin{eqnarray}
    \mathcal{Z}(\lambda,t) &=& {\mathcal{N}} \, \det_{j,t}\big[\overline{{D}}_{\lambda}\big],
    \label{CGF-first}
    \end{eqnarray}
where $\overline{D}_{\lambda}$ in Eq.~\eqref{CGF-first} is discretized in time.
The symbol $j$ corresponds to the system’s degree of freedom and $t$ is the discretized time \cite{kamenev2011field, Bijay_detailed_FCS,Katha_2025}. The normalization constant ${\mathcal{N}}$ can be fixed by demanding $ \mathcal{Z}(0,t)=1$. The normalized MGF in Eq.~\eqref{CGF-first} therefore becomes
\begin{equation}
\mathcal{Z}(\lambda,t) =  \det_{j,t}\big[\overline{{D}}^{-1} \overline{{D}}_{\lambda}\big]= \det_{j,t} \big[\mathbb{I}+ \overline{{D}}^{-1} \overline{{\Sigma}}_{\lambda} \big],
\end{equation}
where we have used the relation in Eq.~\eqref{Dlambda} and $\mathbb{I}$ has a structure akin to Eq.~\eqref{D_0matrix} i.e.,  
${\mathbb I}=
\begin{pmatrix}
I & 0\\
0 & I
\end{pmatrix}
$
with each $I$ being a $L\times L$ matrix. Consequently, the CGF can be expressed as 
\begin{eqnarray}
\chi(\lambda,t) &\equiv& \ln {\cal Z}(\lambda,t)= \mathrm{Tr}_{j,t} \ln \big[\mathbb{I} + \overline{{G}} \, \overline{{\Sigma}}_{\lambda}\big].
\label{CGF-general}
\end{eqnarray}
The matrix $\overline{{G}}$ in Eq.~\eqref{CGF-general} is the discrete version of different components of $G(\tau,\tau')$ [$G_{++}, G_{+-}, G_{-+}, G_{--}$] and can be obtained following Eq.~\eqref{integro-diff}.
It is important to note that the matrix
$\overline{\Sigma}_{\lambda}$, given in Eq.~\eqref{Sigma_lambda}, contains the matrix $P_1$ which has a single nonzero $(1,1)$
element. Consequently,
despite the global nature of the particle-number counting problem,
the CGF in Eq.~\eqref{CGF-general} depends only on the local
Green's function associated with the injection site. The CGF Eq.~\eqref{CGF-general} is one of the central results of this paper. The CGF is valid for arbitrary one-dimensional non-interacting lattices and for arbitrary time instances.

One can rewrite Eq.~\eqref{CGF-general} in a more elegant form by performing an orthogonal Keldysh rotation \cite{kamenev2011field,Bijay_detailed_FCS,rammer2007quantum}
${\cal O}=\frac{1}{\sqrt{2}}
\begin{pmatrix}
I & I\\
I & -I
\end{pmatrix}$ with ${\cal O}^{2}=\mathbb{I}$,
and express the CGF as
\begin{align}
\chi(\lambda,t)
&=
\mathrm{Tr}_{j,t}
\ln\!\left[
\mathbb{I}
+\breve{{G}}\,
\breve{{\Sigma}}_{\lambda}
\right]
\nonumber\\
&= \sum_{n=1}^{\infty} \frac{(-1)^{n+1}}{n} \, \mathrm{Tr}_{j,t} \big[ \big(\breve{{G}}\,
\breve{{\Sigma}}_{\lambda}\big)^n\big].
\label{CGF-general-series}
\end{align}
where the Keldysh-rotated Green's function matrix is given by
(we suppress the time arguments)
\begin{equation}
\breve{G}
=
{\cal O}\,\overline{G}\,{\cal O}^{T}
=
\begin{pmatrix}
0 & G^{A}\\
G^{R} & G^{K}
\end{pmatrix},
\end{equation}
where $G^{(R,A,K)}$ are the retarded ($R$), advanced ($A$), and the Keldysh ($K$) components, respectively. The counting-field-dependent matrix $\overline{\Sigma}_{\lambda}$ transforms as
\begin{equation}
\breve{\Sigma}_{\lambda}
=
{\cal O}\,\overline{\Sigma}_{\lambda}\,{\cal O}^{T}
=
\begin{pmatrix}
a_{\lambda} & -a_{\lambda}\\
a_{\lambda} & -a_{\lambda}
\end{pmatrix},
\end{equation}
where
$a_{\lambda}(t_1,t_2)
=
-i\frac{\Gamma}{2}
\left(e^{i\lambda}-1\right)
P_1 \,\delta(t_1-t_2)$.
In the continuous-time limit, the $n$-th term in the series in Eq.~\eqref{CGF-general-series} takes the form
\begin{eqnarray} && \!\!\!\!\!\mathrm{Tr}_{j,t}
\left[\big(
\breve{{G}}
\breve{{\Sigma}}_{\lambda}\big)^n
\right] =
\int_{t_0}^{t} d\tau_1
\int_{t_0}^{t} d\tau_2 \int_{t_0}^{t} d\tau_3 \cdots \int_{t_0}^{t} d\tau_n \times \nonumber \\
&& \!\!\!\!\mathrm{Tr}_{j}
\left[
\breve{G}(\tau_1,\tau_2)
\breve{\Sigma}_{\lambda}(\tau_2,\tau_3) \breve{G}(\tau_3,\tau_4) \cdots  \breve{\Sigma}_{\lambda}(\tau_n,\tau_1) 
\right].
\label{eq:order-n}
\end{eqnarray}
Following the set of equations Eq.~\eqref{D0++}, Eq.~\eqref{D0+-}, Eq.~\eqref{D0-+}, Eq.~\eqref{D0--} and Eq.~\eqref{integro-diff}, the retarded and advanced Green's functions satisfy the following differential equations
\begin{eqnarray}
\label{Gr-diff}
\Big[i \partial_{\tau} I - h + i \frac{\Gamma}{2} P_1 \Big] {G}^{R}(\tau,\tau') &=&  I \, \delta (\tau-\tau'), \\
\Big[i \partial_{\tau} I - h - i \frac{\Gamma}{2} P_1 \Big] {G}^{A}(\tau,\tau') &=& I \, \delta (\tau-\tau'),
\label{Ga-diff}
\end{eqnarray}
and the Keldysh component is given as \cite{Bijay_detailed_FCS}
\begin{equation}
\label{Gk-diff}
G^{K}(\tau,\tau')= g^K(\tau,\tau')+ i \Gamma \int_{t_0}^{t} d\tau_1 \, G^R(\tau,\tau_1)  \, P_1 \, G^A(\tau_1,\tau'),
\end{equation}
where $g^K(\tau,\tau')$ is the Keldysh component in the absence of the injection term.  Equivalently, these Green's functions can be identified in terms
of the two-time correlation functions. For details please see the supplemental material ~\cite{supp}.
In what follows, we use Eq.~\eqref{CGF-general-series} and obtain both short and long-time limit expressions for the CGF. Note that the probability distribution $p(N,t)$ can be obtained by performing inverse Fourier transformation as $p(N,t)= \int_{0}^{2 \pi} \, d\lambda/ 2 \pi  \, {\cal Z}(\lambda,t) \, \, e^{-i \lambda N}$.

\begin{figure}
\includegraphics[width=0.85\columnwidth]{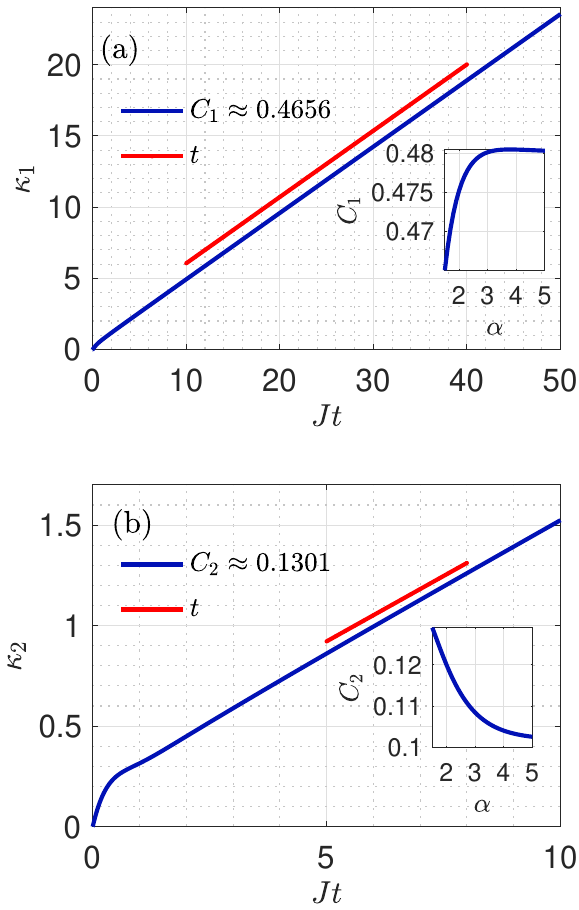}
\caption{Plots for quantum dynamics of particle number growth for first two cumulants: (a) $\kappa_1$, (b) $\kappa_2$ for long-range lattice model, described by the Hamiltonian in Eq.~\eqref{eq:ham} with $\alpha=1.5$. Similar to Fig.~\eqref{fig:tight-binding} asymptotic linear growth is clearly seen and the corresponding slopes match with analytical predictions given in Eq.~\eqref{CGF-long-time}. The plots in insets of (a) and (b) show the dependence of the asymptotic growth rates $C_1$ and $C_2$, respectively, on long-range hopping exponent $\alpha$. The approach to the short range limit is clearly seen in these insets. The parameters used in the simulation are $L=600$, and $\Gamma=1$.}
\label{fig:long-range}
\end{figure}

\vspace{0.2cm} 
\textit{CGF in the short-time limit.--}
We first consider the CGF in the short time limit. As the CGF in Eq.~\eqref{CGF-general-series} is valid for arbitrary time $t$, in the short-time limit (compared to all other dynamical timescales such as $1/\Gamma$, $1/J$ with $J$ being the lattice hopping) we can approximate Eq.~\eqref{CGF-general-series} by considering only the first term in the log expansion. The analytical expression for the CGF then reduces to \cite{supp}
\begin{equation}
\chi(\lambda, T) = \Gamma \, T \, \big(e^{i \lambda}-1\big) + O(t^2),
\label{short}
\end{equation}
with $T=t-t_0$. As a result, all cumulants initially scales linearly with time with identical slope $\kappa_n(t) = \Gamma \,T +O(T^2)$ and the value of the slope is fixed by the injection strength $\Gamma$. Physically, at such early times the injection process has not yet been affected by Pauli exclusion and propagation into the lattice and hence  the dynamics is limited to the local site only. Injection therefore behaves as independent Poisson process. Note that, one can also arrive at this result by making a short time expansion of the dressed generator $e^{L_{\lambda}(t-t_0)} \sim  1 + L_{\lambda} \, (t-t_0)$, with $L_{\lambda}$ given in Eq.~\eqref{GKSL-injection-shifted}.

\vspace{0.2cm}
\textit{CGF in the long-time limit.--}
We next discuss the long-time limit of the CGF following Eq.~\eqref{CGF-general-series}.  An important point to note here is the order in which the thermodynamic limit and the long-time limits are taken. We first take the thermodynamic limit ($L \to \infty$) so that particles injected at the left end can propagate away from the injected site without encountering finite size effect.  For a lattice with delocalized single-particle states, the local Green's functions associated with the injection site lose memory of the initial preparation at sufficiently long times and obey time-translation invariance. 
Under this assumption, the CGF in the long-time limit can be obtained as [see Supplemental material \cite{supp} for the details of the derivation]
\begin{widetext}
\begin{equation}
\chi(\lambda,t \to \infty) = T \int_{-\infty}^{\infty} \frac{d\omega}{2\pi} \ln \Big[ 1 + \Gamma \, G_{11}^{R}[\omega]\, \Gamma_{\rm lat}[\omega]\, G_{11}^A[\omega] \big(e^{i \lambda} -1 \big) \Big],
\label{CGF-long-time}
\end{equation}
\end{widetext}
where $T=t-t_0$ is assumed to be large, $G_{11}^R[\omega] = \big[\omega + i \Gamma/2 - \Sigma^{R}_{\rm lat}[\omega]\big]^{-1}$ with $\Sigma^{R}_{\rm lat}[\omega]$ being the self-energy associated with semi-infinite lattice without the injection site, $G_{11}^A[\omega]= \big[G_{11}^R[\omega]\big]^{*}$ and $\Gamma_{\rm lat}[\omega]= i \big[\Sigma^R_{\rm lat}[\omega] - \Sigma^A_{\rm lat}[\omega]\big]$ is the spectral density.  Eq.~\eqref{CGF-long-time} is similar to the Levitov-Lesovik formula, derived for boundary driven non-interacting fermionic setup, in the steady-state \cite{Levitov1993,Klich2003,KS07,Dhar2007,Bijay_detailed_FCS}.  Following Eq.~\eqref{CGF-long-time} we conclude that all cumulants in the long-time limit scale linearly with time with slope expressed in terms of the local Green function corresponding to the injected site and the self-energy of the remaining semi-infinite lattice. The expressions for the first three scaled cumulants are given as 

\begin{eqnarray}
\frac{\kappa_1}{T} &=& \int_{-\infty}^{\infty} \frac{d\omega}{2\pi} \, \,  T[\omega], \\
 \frac{\kappa_2}{T} &=& \int_{-\infty}^{\infty} \frac{d\omega}{2\pi} \, \,  T[\omega] \,\, \big(1- T[\omega]\big),\\
\frac{\kappa_3}{T} &=& \int_{-\infty}^{\infty} \frac{d\omega}{2\pi} \, \, T[\omega] \,\, \big(1- T[\omega]\big)\, \big(1- 2 T[\omega]\big),
\end{eqnarray}
where $T[\omega]=\Gamma \, G_{11}^{R}[\omega]\, \Gamma_{\rm lat}[\omega]\, G_{11}^A[\omega]$. It is important to note that the emergence of a time-translational invariance regime relies on
the loss of local memory of the initial preparation. This is typically expected
for clean lattice models where the single-particle states relevant to the
injection site are delocalized and transient contributions dephase in the
thermodynamic limit. Such an assumption may, however, break down when the
underlying non-interacting Hamiltonian possesses localized or bound states
having finite overlap with the injection site. Such states can retain local
memory for long times and give rise to non-decaying or slowly decaying
contributions to the local dynamics. Examples include disordered or
quasiperiodic lattices in localized regimes, as well as systems supporting
impurity- or edge-localized states \cite{PhysRevB.73.085119}. In such cases, the approach to a stationary
local description can be absent or extremely slow, and consequently the
simple extensive-in-time form of the CGF in
Eq.~\eqref{CGF-long-time} need not hold
[see Supplemental Material~\cite{supp} for details]. In particular, the initial-condition-dependent contribution to the local Keldysh Green's function can be absent or may decay extremely slowly.

\vspace{0.2em}
\textit{Numerical Results.--} We now present numerical results. We focus on a class of non-interacting lattice setups with Hamiltonian 
\begin{equation}
H= \sum_{\substack{i,j=1\\ i\neq j}}^{L}
J \left(
\frac{1}{|i-j|^{\alpha}} c_i^{\dagger} c_j
+ \mathrm{h.c.}
\right)\,,
\label{eq:ham}
\end{equation}
where $J$ is the hopping amplitude, $\alpha$ is the long-range hopping exponent and $i,j$ represent the lattice sites. The limit $\alpha \to \infty$ corresponds to nearest neighbour tight-binding lattice model. In Fig.~\ref{fig:tight-binding}(a)-(c), we present results for the first three cumulants: $\kappa_1= \mathrm{Tr}(C),\kappa_2=\mathrm{Tr}\left(C-C^2\right)$ and $\kappa_3=\mathrm{Tr}\left(C-3C^2+2C^3\right)$,
where the cumulants are obtained by using Eq.~\eqref{eq:kappa} and Eq.~\eqref{CGF}. For simplicity, we have omitted the $t$ dependence.  We find that in the early time ($t \ll 1/J$), the growth of all cumulants are linear and with the same rate $\Gamma$. This is also clearly seen in the upper inset of Fig.~\ref{fig:tight-binding}(b) and (c). In the long-time limit also, all the three cumulants scale linearly with time i.e., $\kappa_n = C_n t$ where $n=1,2,3$ represents the index of the cumulants. However the slopes $C_n$ are different for different cumulants and matches excellently with analytical computation via Schwinger Keldysh method, as given in Eq.~\eqref{CGF-long-time}. The lower inset of Fig.~\ref{fig:tight-binding}(a)-(c), shows the dependence of the $C_n$ as a function of the injection strength $\Gamma$. We see a clear non-monotonic behavior with $\Gamma$ for all three cumulants.

In Fig.~\ref{fig:long-range}, we plot the first two cumulants for particle number statistics for long-range lattice model, as given in Eq.~\eqref{eq:ham} for $\alpha=1.5$. We observe that similar to the tight-biding case, the cumulants grow linearly in the long-time limit and the corresponding slopes match with the analytical predictions, as given in Eq.~\eqref{CGF-long-time}. We further plot (insets) the dependence of the asymptotic growth rates $C_1$ and $C_2$, on the long-range hopping exponent $\alpha$. Interestingly, we observe opposite trend in the behaviour of $C_1$ and $C_2$ as a function of $\alpha$. The approach to the short range limit is also clearly seen in these insets. 

\vspace{0.2cm}
\textit{Summary and outlook.--}
We investigated the FCS of the total particle number growth in a generic free-fermionic lattice driven by a localized source. Using the Schwinger-Keldysh path integral formalism within the GKSL framework, we derived an exact expression for the CGF that depends on the local Green's function corresponding to the injected site and a self-energy due to the remaining part of the lattice. This enabled us to obtain an explicit analytical expression for the long-time CGF and to demonstrate that, for clean short- and long-range lattices hosting delocalized single-particle states, all cumulants grow linearly with time in the asymptotic regime. Our analytical predictions are corroborated by direct numerical simulations for both short- and long-range lattice models. Interestingly, we demonstrate that lattice models that do not support delocalized states, can exhibit growth rates very different from linear (for e.g. $\ln t$).

Our results establish a general framework for investigating quantum fluctuations in driven free-fermionic systems with localized gain. The formalism naturally extends to multiple gain and loss channels as well as to higher-dimensional lattices. An interesting future direction is the bosonic counterpart, where localized gain is known to induce nonequilibrium dynamical phase transitions in particle-number growth~\cite{Krapivsky_2020}, potentially leading to qualitatively different counting statistics. An even more challenging and exciting direction is to generalize the present framework to interacting many-body systems.

\vspace{0.2cm}
\textit{Acknowledgements.--}   BKA acknowledges the CRG grant No. CRG/2023/003377 from ANRF, Government of India. BKA thanks Katha Ganguly for useful discussions. MK acknowledges support from the Department of Atomic Energy, Government of India, under Project No. RTI4001.
MK thanks the hospitality of Laboratoire de Physique Théorique et Modèles Statistiques (LPTMS), University Paris-Saclay and Collège de France, PSL Research University where a major part of the work took place.

\let\oldaddcontentsline\addcontentsline
\renewcommand{\addcontentsline}[3]{}
\bibliography{references}
\let\addcontentsline\oldaddcontentsline%

\clearpage

\newpage
\appendix
\onecolumngrid
\setcounter{figure}{0}
\setcounter{equation}{0}
\setcounter{page}{1}
\setcounter{figure}{0}
\renewcommand{\thefigure}{S\arabic{figure}}
\numberwithin{equation}{section}
\thispagestyle{empty}

\begin{center}
    {\bfseries\large
    \underline{Supplemental Material} \\[1ex]
    ``Quantum Dynamics of Full Counting Statistics in Fermionic Lattices with Localized Gain''}\\
    


\end{center}

\tableofcontents

\section{Derivation of Eq.~\ref{alternate-GF} and Eq.~\ref{GKSL-injection-shifted} of the main text}
\label{alternate}
In this section, we derive the alternative expression of the MGF, as given in Eq.~\eqref{alternate-GF}. For simplicity we take here the initial time $t_0=0$. 
We start from the definition of the MGF and write,
\begin{equation}
{\cal Z}(\lambda,t) = \langle e^{i \lambda N(t)} \rangle. 
\end{equation}
Switching from the Heisenberg to the Schr\"odinger picture, we obtain 
\begin{equation}
{\cal Z}(\lambda,t)
= {\rm Tr}\Big[e^{i \lambda N} \rho(t)\Big] = {\rm Tr}\Big[e^{i \lambda N} e^{Lt}\rho(0)\Big] 
= {\rm Tr}\Big[e^{i \lambda N} e^{Lt} \big[e^{-i \lambda N}\rho(0)\big]\Big],
\label{end-eq}
\end{equation}
where in the last expression we introduce $e^{-i \lambda N(0)}$ by assuming that the initial state $\rho(0)$ is a vacuum state. In that case as $N \rho(0)= 0$ and we get $e^{-i \lambda N}\rho(0)=\rho(0)$.  To make the action of the counting-field transformation explicit,
we introduce the left-multiplication superoperator
\begin{equation}
{\cal S}_{\lambda}X
=
e^{i\lambda N}X,
\qquad
{\cal S}_{\lambda}^{-1}X
=
e^{-i\lambda N}X.
\end{equation}
The tilted Liouvillian is then defined as
\begin{equation}
{L}_{\lambda}
=
{\cal S}_{\lambda}
{L}
{\cal S}_{\lambda}^{-1}
\label{L-lambda-tilted}
\end{equation}
and as a result
\begin{equation}
e^{{L}_{\lambda}t}
=
{\cal S}_{\lambda}
e^{{L}t}
{\cal S}_{\lambda}^{-1},
\end{equation}
and we can expres the MGF in Eq.~\eqref{end-eq} as
\begin{equation}
{\cal Z}(\lambda,t)= {\rm Tr}\big[e^{L_{\lambda} t} \rho(0)],
\label{GF}
\end{equation}
Given the GKSL equation with local injection
\begin{equation}
\frac{d \rho}{dt}=-i \big[H, \rho \big] + \Gamma \big[c_1^{\dagger} \rho c_1 - \frac{1}{2} \big\{c_1 c_1^{\dagger}, \rho\big \}] \equiv L \, \rho,
\label{GKSL-injection-app}
\end{equation}
let us decompose the total Liouvillian $L$ as  a sum of unitary $L_u$ and a dissipative part $L_d$, i.e., $L=L_u + L_d$ where $L_u \, * = - i [H, *]$ and $L_d \, * = \gamma \big[c_1^{\dagger} * c_1 - \frac{1}{2} \big\{c_1 c_1^{\dagger}, * \big \}]$. As a result, $L_{\lambda}$ in Eq.~\eqref{L-lambda-tilted} can be written as $L_{\lambda}= L_{u}^{\lambda} + L_{d}^{\lambda}$.  Now for a number conserving Hamiltonian $[H,N]=0$, $L_u^{\lambda}=L_u$, i.e., independent of $\lambda$. However, this is not the case for $L^d_{\lambda}$. We write 
\begin{eqnarray}
L_{d}^{\lambda} \, \rho &=& \Gamma \Big[\big(c^{\lambda}_1)^{\dagger} \,  \rho \, c_1 - \frac{1}{2} \Big\{c_1^{\lambda} \big(c^{\lambda}_1)^{\dagger}, \rho\Big \}\Big],
\end{eqnarray}
where $\big(c^{\lambda}_1\big)^{\dagger}= e^{i \lambda N(0)} \, c_1^{\dagger}\,e^{-i \lambda N(0)} = e^{i \lambda } \, c_1^{\dagger}$ and $\big(c^{\lambda}_1\big)= e^{i \lambda N(0)} \, c_1 \,e^{-i \lambda N(0)} = e^{-i \lambda } \, c_1 $. As a result, only the first term in $L_{d}^{\lambda}$ carries $\lambda$ dependence and it appears as a phase. The final expression for $L_{\lambda}$ is given as 
\begin{equation}
L_{\lambda} * =-i \big[H, * \big] + \Gamma \Big[e^{i \lambda} \, c_1^{\dagger} * c_1 - \frac{1}{2} \big\{c_1 c_1^{\dagger}, *\big \}\Big],
\label{GKSL-injection-shifted-supp}
\end{equation}
which matches with Eq.~\eqref{GKSL-injection-shifted} of the main text.

\section{Derivation of the action in Eq.~\ref{action1} following Schwinger-Keldysh path-integral formalism}
\label{Keldysh}
In this section, we derive the action given in Eq.~(\ref{action1}) of the main text starting from the MGF, as given in Eq.~\eqref{alternate-GF}. We follow the Schwinger-Keldysh path-integral approach to derive the action \cite{Katha_2025}.
We start with the definition of the MGF and write
\begin{equation}
{\cal Z}(\lambda,t)= {\rm Tr}\big[e^{L_{\lambda}(t-t_0)} \rho(t_0)\big]= {\rm Tr}\big[\rho_{\lambda}(t)\big].
\label{GF-app}
\end{equation}
We discretize the time window $t-t_0=N \delta t$ such that $N \to \infty$ and $\delta t \to 0$, keeping $N \delta t$ finite. We denote $t_n= n\,\delta t$ and write the single step evolution as
\begin{equation}
\rho_{\lambda}(t_{n+1}) = e^{L_{\lambda} \delta t} \, \rho_{\lambda}(t_{n}). 
\label{liou-dyn}
\end{equation}
Considering $\delta t \rightarrow 0$ limit, we expand the exponential in Eq.~\eqref{liou-dyn} and write 
\begin{equation}
\rho_{\lambda}(t_{n+1}) = \big[\mathbb{I} + \delta t \, L_{\lambda}\big] \, \rho_{\lambda}(t_{n}). 
\label{time-density-evolve}
\end{equation}
We now use of path integral approach and introduce the Grassmann variables ~\cite{kamenev2011field,Sieberer_2016}, which are the coherent basis set for the fermions. These variables satisfy 
\begin{align}    c_{i}|\psi_i\rangle=\psi_{i}|\psi_i\ra ,\quad i = 1,2,\dots,L
\end{align}
where $\psi_i$ and $|\psi_i\rangle$ are the eigenvalue and the corresponding eigenvector for the annihilation operator $c_{i}$. As $c_i$ are the fermionic operators, the eigenvalues $\psi_i$ satisfies the following anticommutation relation,
\begin{align}
    \psi_i \psi_j+\psi_j\psi_i=0,\quad \psi_i^2=0,\quad \psi_i^{*}\psi_j+\psi_j^{*}\psi_i=\delta_{ij}.
\end{align}
The coherent basis set is overcomplete and follow the completeness relation 
\begin{align}
    \int {d\psi^{*}\, d\psi}\,e^{-\Bar{\psi}\psi}\,|\psi\rangle\la\psi|=\mathbb{I}, \label{eq:identity_coherent_supp}
\end{align}
where $|\psi \rangle$ is a column vector containing $|\psi_i\rangle$ for all the lattice sites. Note that, these basis vectors are  not orthogonal and satisfies $\la \psi_i|\phi_j\rangle=e^{\psi^{*}_i\phi_j}$. 
We now write $\rho_{\lambda}(t_n)$ in Eq.~\eqref{time-density-evolve} by introducing the Grassmann fields $|\psi_i \rangle$ and use the completeness relation in Eq.~\eqref{eq:identity_coherent_supp}. We then obtain \cite{Katha_2025}
\begin{equation}
\rho_{\lambda}(t_n)= \int D \psi_{n}^{+} \int D \psi_{n}^{-}\,  e^{-\psi_{n +}^{*} \psi_{n +}}\,  e^{-\psi_{n -}^{*} \psi_{n -}}\, |\psi_{n}^{+}\rangle \, \langle \psi_{n}^{+}| \rho_{\lambda}(t_n)  |\psi_{n}^{-}\rangle \, \langle \psi_{n}^{-}|.
\label{basis-exp1}
\end{equation}
Following Eq.~\eqref{time-density-evolve}, we compute the matrix element 
\begin{equation}
\langle \psi_{n+1}^{+}|\rho_{\lambda}(t_{n+1})|\psi_{n+1}^{-}\rangle = \langle \psi_{n+1}^{+}|\rho_{\lambda}(t_{n})|\psi_{n+1}^{-}\rangle + \delta t\, \langle \psi_{n+1}^{+}| L_{\lambda} \rho_{\lambda}(t_n)|\psi_{n+1}^{-}\rangle  + O(\delta t^2)
\end{equation}
which we further express in terms of Eq.~\eqref{basis-exp1} as
\begin{eqnarray}
&&\langle \psi_{n+1}^{+}|\rho_{\lambda}(t_{n+1})|\psi_{n+1}^{-}\rangle = \int D \psi_{n}^{+} \int D \psi_{n}^{-} \, e^{-\psi_{n +}^{*} \psi_{n +}}\,  e^{-\psi_{n -}^{*} \psi_{n -}}\, \langle \psi_{n+1}^{+}|\psi_{n}^{+} \rangle \langle \psi_{n}^{+} | \rho_{\lambda}(t_{n})|\psi_{n}^{-}\rangle \langle \psi_{n}^{-} | \psi_{n+1}^{-}\rangle  \nonumber \\
 &&+ \delta t\, \int D \psi_{n}^{+} \int D \psi_{n}^{-}  e^{-\psi_{n +}^{*} \psi_{n +}}\,  e^{-\psi_{n -}^{*} \psi_{n -}}\, \langle \psi_{n+1}^{+}| L_{\lambda} \Big[ |\psi_{n}^{+}\rangle \, \langle \psi_{n}^{-}| \Big]  |\psi_{n+1}^{-}\rangle \, \langle \psi_{n}^{+} | \rho_{\lambda}(t_n) |\psi_{n}^{-}\rangle. \,  \,\,
 \label{details-rho}
\end{eqnarray}
Recall that the action of the dressed Liouvillian ${L}_{\lambda}$ on the density matrix $\rho$ is,
\begin{align}
    {L}_\lambda\,\rho=-i\,H\,\rho + i\, \rho\, H+  \Gamma \Big[e^{i \lambda} \, c_1^{\dagger} \, \rho \, c_1 - \frac{1}{2} \big\{c_1 c_1^{\dagger}, \rho \big \}\Big].
    \label{liou_boundry}
\end{align}
Therefore, following Eq.~\eqref{details-rho}, we need to calculate $\la \psi_{(n+1)+}|{L}_{\lambda}\big[|\psi_{n+}\ra\la\psi_{n-}|\big]|\psi_{(n+1)-}\ra$, which we write as follows:
\begin{align}
\la\psi_{(n+1)^{-}}|{L}_\lambda\,\big[|\psi_{n+}\ra\la\psi_{n-}|\big]|\psi_{(n+1)-}\ra&=f_\lambda(\psi^{*}_{(n+1)+},\psi_{n+},\psi^{*}_{n-},\psi_{(n+1)-})\, \,\,\la\psi_{(n+1)+}|\psi_{n+}\ra\,\,\la\psi_{n-}|\psi_{(n+1)-}\ra.\label{liouvel_coherent_supp}
\end{align}
We can express $f_\lambda(\psi^{*}_{(n+1)+},\psi_{n+},\psi^{*}_{n-},\psi_{(n+1)-})$ as the sum of two terms. The first term is the contribution coming from the unitary part, $f_u(\psi^{*}_{(n+1)+},\psi_{n+},\psi^{*}_{n-},\psi_{(n+1)-})$. We write 
\begin{align}
    -i\Big(\la\psi_{(n+1)+}| &H|\psi_{n+}\ra\la\psi_{n-}|\psi_{(n+1)-}\ra - \la\psi_{(n+1)+}|\psi_{n+}\ra\la\psi_{n-}| H|\psi_{(n+1)-}\ra\Big)\nonumber\\&=-i\sum_{i,j=1}^{N}h_{ij}\Big[\psi_{(n+1)+}^{i*}\psi^{j}_{n+}-\psi_{n-}^{i*}\psi^{j}_{(n+1)-}\Big] \,\la\psi_{(n+1)+}|\psi_{n+}\ra\la\psi_{n-}|\psi_{(n+1)-}\ra,\label{hamiltonian1_coh_supp}
\end{align}
and identify the unitary part contribution $f_u$ as,
\begin{align}
f_u(\psi^{*}_{(n+1)+},\psi_{n+},\psi^{*}_{n-},\psi_{(n+1)-})=-i\sum_{i,j=1}^{L}h_{ij}\Big[\psi_{(n+1)+}^{i*}\psi^{j}_{n+}-\psi_{n-}^{i*}\psi^{j}_{(n+1)-}\Big]. \label{fH1_supp_boundary}
\end{align}
Similarly,  the for the dissipative part, $f^{\lambda}_d(\psi^{*}_{(n+1)+},\psi_{n+},\psi^{*}_{n-},\psi_{(n+1)-})$ 
we obtain
\begin{align} f^{\lambda}_d(\psi^{*}_{(n+1)+},\psi_{n+},\psi^{*}_{n-},\psi_{(n+1)-})&=\Gamma \Big [e^{i\lambda}\psi^{1*}_{(n+1)+}\psi_{(n+1)-}^{1}+ \frac{1}{2} \psi^{1*}_{(n+1)+}\psi_{n+}^{1}+ \frac{1}{2} \psi^{1*}_{n-}\psi_{(n+1)-}^{1}\Big].
    \label{fd_supp}
\end{align}
Combining Eq.~\eqref{fH1_supp_boundary} and Eq.~\eqref{fd_supp}, one can obtain the full expression of $\la\psi_{(n+1)+}|{L}_\lambda\big[|\psi_{n+}\ra\la\psi_{n+}|\big]|\psi_{(n+1)-}\ra$. Using that, finally we write $\la\psi_{(n+1)+}|\rho^{\lambda}_{n+1}|\psi_{(n+1)-}\ra$ as, 
\begin{align}
\la\psi_{(n+1)+}|\rho^{\lambda}_{n+1}|\psi_{(n+1)-}\ra=\int {d\psi^{*}_{n+}d\psi_{n+}}{d\psi^{*}_{n-}d\psi_{n-}}&e^{-(\psi^{*}_{n+}\psi_{n+}+\psi^{*}_{n-}\psi_{n-})}e^{(\psi_{(n+1)+}^{*}\psi_{n+}+\psi_{n-}^{*}\psi_{(n+1)-})}\nonumber\\& \times 
 \Big[1+\delta t f_{\lambda}(\psi^{*}_{(n+1)+},\psi_{n+},\psi^{*}_{n-},\psi_{(n+1)-})\Big]\la\psi_{n+}|\rho^{\lambda}_n|\psi_{n-}\ra,\label{rho_n1_coh}
\end{align}
where $f_\lambda=f_u+f_d^{\lambda}$.  We next take the continuum time limit by considering $\delta t \rightarrow 0 $ and write $\big[\psi_{(n+1)\pm}-\psi_{n\pm}\big] (\delta t)^{-1}=\partial_t\psi_{n \pm}$. 
Thus the counting field dressed density matrix at time $t$ in terms of the Grassmann variables takes the form,
\begin{align}
    \rho_{\lambda}(t)=\int \prod_{n=1}^{\frac{t-t_0}{\delta t}}{d\psi_{n+}^*d\psi_{n+}}{d\psi_{n-}^*d\psi_{n-}}\,\,&e^{\delta_t[-\psi_{n+}\partial_t\psi^{*}_{n+}+\psi^{*}_{n-}\partial_t\psi_{n-}+f_\lambda(\psi^{*}_{(n+1)+},\psi_{n+},\psi^{*}_{n-},\psi_{(n+1)-})]}\nonumber\\& \times |\psi_{(t-t_0)/\delta_t+}\ra\la\psi_{(t-t_0)/\delta_t-}| \,\la\psi_{0+}|\rho(t_0)|\psi_{0-}\ra \label{eq:rho_lam_t}
\end{align}
Using integration by parts for the Grassmann variables, we finally identify the action as,
\begin{align}
    S(\lambda) = \int_{t_0}^{t_f} dt  \,\big[\psi^{*}_{+} \big(i\partial_t \big) \psi_{+}-\psi^{*}_{-} \big(i\partial_t\big) \psi_{-}]-if_{\lambda}(\psi^{*}_{+},\psi_{+},\psi^{*}_{-},\psi_{-})\big], 
    \label{action_supp}
\end{align}
where $f_{\lambda}(\psi^{*}_{+},\psi_{+},\psi^{*}_{-},\psi_{-})$ is given as follows
\begin{align}
    f_{\lambda}(\psi^{*}_{+},\psi_{+},\psi^{*}_{-},\psi_{-})=&-i\sum_{i,j=1}^{L}h_{ij} [\psi_{i+}^*\psi_{j+}-\psi^*_{i-}\psi_{j-}]+ \Gamma \big(e^{i\lambda}\psi^{*}_{1,+}\psi_{1,-}+\frac{1}{2} \psi^{*}_{1,+}\psi_{1,+}+ \frac{1}{2}  \psi^{*}_{1,-}\psi_{1,-}\big)
    \label{supp-f-lambda}
\end{align}
Eq.~\eqref{action_supp} along with Eq.~\eqref{supp-f-lambda} is the appropriate action, as given in the main text in Eq.~\eqref{action1}, Eq.~\eqref{fH}, and Eq.~\eqref{dissipative}.

\section{Derivation of the long-time CGF given in Eq.~\ref{CGF-long-time}}
In this section, we derive the long-time limit expression for the CGF as given in  Eq.~\eqref{CGF-long-time} of the main text.
The normalized CGF is expressed as (Eq.~\eqref{CGF-general-series} of the main text) 
\begin{align}
\chi(\lambda,t)
=
\mathrm{Tr}_{j,t}
\ln\!\Big[
\mathbb{I}
+\breve{{G}}\,
\breve{{\Sigma}}_{\lambda}
\Big]= \sum_{n=1}^{\infty} \frac{(-1)^{n+1}}{n} \, \mathrm{Tr}_{j,t} \big[ \big(\breve{{G}}\,
\breve{{\Sigma}}_{\lambda}\big)^n\big].
\end{align}
In the continuous time notation, the $n$-th order term in the series can be written as 
\begin{eqnarray}  \!\!\mathrm{Tr}_{j,t}
\left[\big(
\breve{{G}}
\breve{{\Sigma}}_{\lambda}\big)^n
\right] =
\int_{t_0}^{t} d\tau_1
\int_{t_0}^{t} d\tau_2 \int_{t_0}^{t} d\tau_3 \cdots \int_{t_0}^{t} d\tau_n  \mathrm{Tr}_{j}
\left[
\breve{G}(\tau_1,\tau_2)
\breve{\Sigma}_{\lambda}(\tau_2,\tau_3) \breve{G}(\tau_3,\tau_4) \cdots  \breve{\Sigma}_{\lambda}(\tau_n,\tau_1) 
\right].
\label{eq:order-n}
\end{eqnarray}
Recall that the Keldysh rotated matrix $\breve{{G}}(t_1,t_2)$ is given as
\begin{eqnarray}
\breve{{G}}(t_1,t_2) 
&=& \begin{pmatrix}
0 & {G}^A (t_1,t_2)\\
{G}^R (t_1,t_2) & {G}^K (t_1,t_2)
\end{pmatrix}
\end{eqnarray}
where $G^{R}$, $G^{A}$, and $G^{K}$ refers to the retarded, advanced, and the Keldysh components, respectively and all the block matrices are of size $L \times L$. In the time-domain the retarded ($R$) and the advanced ($A$) Green's functions satisfy the differential equations given in Eq.~\eqref{Gr-diff} and Eq.~\eqref{Ga-diff}. The Keldysh component satisfies Eq.~\eqref{Gk-diff}. Equivalently, these Green's functions can be identified in terms of the two-time correlation functions, given as  \cite{rammer2007quantum, kamenev2011field}
\begin{align}
\label{Gr-corr}
    \big[G^R(\tau,\tau')\big]_{ij}
&=
-i\theta(\tau-\tau')
\big\langle
\big\{
c_i(\tau),c_j^\dagger(\tau')
\big\}
\big\rangle ,
\\
\label{Ga-corr}
\big[G^A(\tau,\tau')\big]_{ij}
&=
+i\theta(\tau'-\tau)
\big\langle
\big\{
c_i(\tau),c_j^\dagger(\tau')
\big\}
\big\rangle ,
\\
\label{Gk-corr}
\big[G^K(\tau,\tau')\big]_{ij}
&=
-i
\big\langle
\big[
c_i(\tau),c_j^\dagger(\tau')
\big]
\big\rangle .
\end{align}
The counting field dependent matrix $\breve{\Sigma}_{\lambda}$, appearing in Eq.~\eqref{eq:order-n} is given as
\begin{eqnarray}
\breve{\Sigma}_{\lambda}(t_1,t_2) 
&=& \begin{pmatrix}
a_{\lambda}(t_1,t_2) & -a_{\lambda} (t_1,t_2)\\
a_{\lambda} (t_1,t_2)& -a_{\lambda}(t_1,t_2)
\end{pmatrix},
\end{eqnarray}
where 
\begin{equation}
    a_{\lambda}(t_1,t_2)=-i \frac{\Gamma}{2} (e^{i \lambda}-1) P_1 \delta(t_1 - t_2).
    \label{a-lambda}
\end{equation} 
Recall that $P_1$ is a $L\times L$ matrix with only the $(1,1)$ element is non-zero with value $1$ and all other elements are zero.  Thus we can write Eq.~\eqref{eq:order-n} for $n=1$ as
\begin{equation}
 \mathrm{Tr}_{j,t} \big[\breve{G} \, \breve{{\Sigma}}_{\lambda}\big]  = \int_{t_0}^{t} dt_1 \, \int_{t_0}^{t} dt_2 \, \mathrm{Tr}_{j}\big[\breve{G}(t_1,t_2)\,\breve{\Sigma}_{\lambda}(t_2,t_1) \big] = \int_{t_0}^{t} dt_1 \, \mathrm{Tr}_{j}\big[\breve{G}(t_1,t_1)\, \breve{\widetilde{\Sigma}}_{\lambda} \big],
 \label{delta}
 \end{equation}
where $\breve{\widetilde{\Sigma}}_{\lambda}$ is now time-independent has entries $\widetilde{a}_{\lambda}$ with  
\begin{equation}
\widetilde{a}_{\lambda}=-i \frac{\Gamma}{2} (e^{i \lambda}-1) P_1.
\end{equation}
We now consider the long-time behaviour of the CGF. An important point is the order in which the thermodynamic limit and the long-time limits are taken. We first take the thermodynamic limit ($L \to \infty$) so that particles injected at the left end can propagate away from the injected site without encountering finite size effect. 
For a lattice with delocalized single-particle states, the local Green's functions associated with the injection site lose memory of the initial preparation at sufficiently long times and approach
a time-translationally invariant form. Thus $\breve{G}(t_1,t_2)$ becomes a function of time difference 
$\breve{G}(t_1-t_2)$ in the long-time limit. Equivalently, the initial time can be taken sufficiently far in the
past, $t_0\rightarrow-\infty$, such that transient contributions
associated with the initial condition do not contribute to the
leading long-time behavior. 
Under the time-translational invariance, the Green's functions can be Fourier
transformed as
\begin{equation}
    \breve{G}(t_1-t_2)
    =
    \int_{-\infty}^{\infty}
    \frac{d\omega}{2\pi}\,
    e^{-i\omega(t_1-t_2)}
    \breve{G}[\omega].
    \label{eq:GF-Fourier}
\end{equation}
To see how the long-time dependence emerges, consider the
$n$-th order contribution in the logarithmic expansion of the CGF,
${\rm Tr}_{j,t}
    \Big[
    \Big(
    \breve{{G}}
    \breve{{\Sigma}}_{\lambda}
    \Big)^n
    \Big]$. 
As in the long-time limit, the local Green's functions depend only on
time differences, one of the time integrations
corresponds to an overall translation of all time arguments and
produces the time window $T=(t-t_0)$, while the remaining integrations
involve only relative times. In the long-time limit one therefore
obtains
\begin{equation}
    {\rm Tr}_{j,t}
    \Big[
    \Big(
    \breve{{G}}\,
    \breve{{\Sigma}}_{\lambda}
    \Big)^n
    \Big]
    \xrightarrow[T\rightarrow\infty]{}
    T
    \int_{-\infty}^{\infty}
    \frac{d\omega}{2\pi}\,
    {\rm Tr}_{j}
    \Big[
    \Big(
    \breve{G}[\omega]
    \breve{\widetilde{\Sigma}}_{\lambda}
    \Big)^n
    \Big].
    \label{eq:nth-order-long-time}
\end{equation}
Using the expansion
\begin{equation}
    \ln(\mathbb{I}+X)
    =
    \sum_{n=1}^{\infty}
    \frac{(-1)^{n+1}}{n}X^n,
\end{equation}
and resumming the series, the CGF in the long-time limit takes
the form
\begin{equation}
    \chi(\lambda,T)
    =
    T
    \int_{-\infty}^{\infty}
    \frac{d\omega}{2\pi}\,
    {\rm Tr}_{j}
    \ln
    \Big[
    \mathbb{I}
    +
    \breve{G}[\omega]
  \breve{\widetilde{\Sigma}}_{\lambda}
    \Big]
    \label{eq:CGF-long-time-general}
\end{equation}
Therefore, the CGF becomes extensive in the time $T$,
It follows immediately that all cumulants grow linearly
with time in the asymptotic regime.
One can simplify the expression in Eq.~\eqref{eq:CGF-long-time-general} further by explicitly multiplying the matrices $\breve{G}[\omega]$ and $\breve{\widetilde{\Sigma}}_{\lambda}$. We obtain 
\begin{eqnarray}
\chi(\lambda,t \to \infty) = T \, \int_{-\infty}^{\infty} \frac{d\omega}{2 \pi} \,  \ln \Big[1 + \frac{i \, \Gamma}{2} (e^{i \lambda} -1 ) \Big[{G}^R_{11}[\omega]- {G}^A_{11}[\omega] +G_{11}^{K}[\omega]\Big]\Big].
\label{CGF-long-time-generic}
\end{eqnarray}
Eq.~\eqref{CGF-long-time-generic} is valid in the long-time limit for generic non-interacting one-dimensional lattice supporting delocalized single particle states.

We now recast Eq.~\eqref{CGF-long-time-generic} into a more
elegant form for generic one-dimensional non-interacting
lattice systems. Since particle injection occurs only at the first site,
in the thermodynamic limit, $L\rightarrow\infty$, we can regard
site $1$ as a local site coupled to the remaining semi-infinite
lattice. The effect of the latter can then be incorporated through
a lattice self-energy. The retarded Green's function at the injection site satisfies the Dyson equation \cite{rammer2007quantum, Bijay_detailed_FCS}
\begin{equation}
G_{11}^R[\omega]
=
G_{0,11}^R[\omega]
+
G_{0,11}^R[\omega]\,
\Sigma_{\rm lat}^R[\omega]\,
G_{11}^R[\omega],
\label{Gr-11}
\end{equation}
where
\begin{equation}
G_{0,11}^R[\omega]
=
\frac{1}{\omega+i\Gamma/2}
\end{equation}
is the retarded Green's function of the injection site
in the presence of the local gain process. We assume here that the onsite energy
of the injection site is zero. Equation~\eqref{Gr-11}
therefore gives
\begin{equation}
G_{11}^R[\omega]
=
\frac{1}{
\omega+i\Gamma/2-\Sigma_{\rm lat}^R[\omega]
},
\label{Gr-11-explicit}
\end{equation}
where $\Sigma_{\rm lat}^R[\omega]$ denotes the retarded self-energy
arising from the coupling of site $1$ to the remaining lattice.
The corresponding advanced Green's function is $G_{11}^A[\omega]
=
\left[G_{11}^R[\omega]\right]^\dagger$.
In the stationary long-time limit, the Keldysh component obeys \cite{rammer2007quantum}
\begin{equation}
G_{11}^K[\omega]
=
G_{11}^R[\omega]
\left[
\Sigma_{\rm inj}^K
+
\Sigma_{\rm lat}^K[\omega]
\right]
G_{11}^A[\omega],
\end{equation}
where for the local particle-injection case,
$\Sigma_{\rm inj}^K=i\Gamma$. Moreover, as the remaining
lattice is viewed as initially empty, we can write  $\Sigma_{\rm lat}^K[\omega]
=
-i\Gamma_{\rm lat}[\omega]$, where
$
\Gamma_{\rm lat}[\omega]
=
i\left[
\Sigma_{\rm lat}^R[\omega]
-
\Sigma_{\rm lat}^A[\omega]
\right].
$
As a result,
\begin{equation}
G_{11}^K[\omega]
=
i\,G_{11}^R[\omega]
\left[
\Gamma-\Gamma_{\rm lat}[\omega]
\right]
G_{11}^A[\omega].
\label{Gk-11}
\end{equation}
It is also useful to note that Eq.~\eqref{Gr-11-explicit} implies
the spectral identity
\begin{equation}
i\left[
G_{11}^R[\omega]-G_{11}^A[\omega]
\right]
=
G_{11}^R[\omega]
\left[
\Gamma+\Gamma_{\rm lat}[\omega]
\right]
G_{11}^A[\omega].
\label{spectral-identity}
\end{equation}
Using Eqs.~\eqref{Gk-11} and \eqref{spectral-identity} in
Eq.~\eqref{CGF-long-time-generic}, the long-time CGF reduces to
\begin{equation}
\chi(\lambda, T \to \infty) = T \int_{-\infty}^{\infty} \frac{d\omega}{2\pi} \ln \Big[ 1 + \Gamma \, G_{11}^{R}[\omega]\, \Gamma_{\rm lat}[\omega]\, G_{11}^A [\omega] \big(e^{i \lambda} -1 \big) \Big],
\label{CGF-long-time-supp}
\end{equation}
which is Eq.~\eqref{CGF-long-time} of the main text.

\section{CGF in the short time limit}
In this section, we derive the expression for the CGF in the short-time limit, as given in Eq.~\eqref{short} of the main text.

\subsection{Short-time CGF following the dressed GKSL equation}
We first derive the short time CGF starting with the dressed GKSL equation. For simplicity we set the initial time $t_0=0$. Recall that the definition of the MGF in Eq.~\eqref{GF} is given as
\begin{equation}
{\cal Z}(\lambda,t)= {\rm Tr}\big[e^{L_{\lambda}t} \rho(0)\big]= {\rm Tr}\big[\rho_{\lambda}(t)\big].
\label{GF-app-2}
\end{equation}
We make a short-time expansion of $e^{L_{\lambda}t} \approx I + L_{\lambda} t$ assuming $t \ll 1/J, 1/\Gamma$,
and calculate ${\cal Z}(\lambda,t)$. We obtain 
\begin{equation}
{\cal Z}(\lambda,t) =  1 + t  \, {\rm Tr}\big[ L_{\lambda} \rho(0)\big] + O(t^2).
\end{equation}
The dressed Liouvillian $L_{\lambda}$ is given as
\begin{equation}
L_{\lambda} * =-i \big[H, * \big] + \Gamma \Big[e^{i \lambda} \, c_1^{\dagger} * c_1 - \frac{1}{2} \big\{c_1 c_1^{\dagger}, *\big \}\Big]
\label{GKSL-injection-shifted-supp-2}
\end{equation}
As a result, we obtain 
\begin{equation}
{\rm Tr}\big[ L_{\lambda} \rho(0)\big] = \Gamma  \, (e^{i \lambda}-1) \, {\rm Tr} \big[c_1 c_1^{\dagger} \rho_0].
\end{equation}
For initial state with empty lattice  ${\rm Tr} \big[c_1 c_1^{\dagger} \rho_0]=1$ and as a result, we obtain the MGF as
\begin{equation}
{\cal Z}(\lambda) =   1 + t\, \Gamma\, (e^{i \lambda}-1) + O(t^2) 
\end{equation}
Therefore the CGF is given as
\begin{equation}
\chi(\lambda, t) = \Gamma \, t \, \big(e^{i \lambda}-1\big) + O(t^2),
\end{equation}
which matches with Eq.~\eqref{short} in the main text. The above results implies that all cumulants, in the short time limit grows linearly with time and with identical slope. The cumulants are given by $\kappa_n(t) = \Gamma\, t + O(t^2)$.

\subsection{Short time CGF following the Green's function approach}
We next obtain the same result from the CGF expression in Eq.~\eqref{CGF-general}. We write the CGF as
\begin{eqnarray}
\chi(\lambda,t) &=& \mathrm{Tr}_{j,t} \ln \big[{I} + \breve{{G}} \, \breve{{\Sigma}}_{\lambda}\big], \nonumber \\
&=& \mathrm{Tr}_{j,t} \big[\breve{{G}} \, \breve{{\Sigma}}_{\lambda}\big] + \frac{1}{2} \mathrm{Tr}_{j,t}  \big[\breve{{G}} \, \breve{{\Sigma}}_{\lambda} \breve{{G}} \, \breve{{\Sigma}}_{\lambda}\big] + \cdots
\label{CGF-general-supp}
\end{eqnarray}
It is easy to see that in the short time limit, only the first term of the log series contribute. As a result, we get (setting the initial time $t_0=0$ for simplicity)
\begin{eqnarray}
\chi(\lambda, t) &\approx&  \frac{i \Gamma}
{2} (e^{i \lambda} - 1) \int_{0}^{t} dt_1 \Big[G_{11}^R(t_1,t_1)-G_{11}^A(t_1,t_1) + G_{11}^K(t_1,t_1)\Big]. 
\end{eqnarray}
Now using the expressions for the Green's functions in terms of the correlators, as given in Eqs.~\eqref{Gr-corr}, \eqref{Ga-corr}, and \eqref{Gk-corr}, we can write $G_{11}^R(t_1,t_1)= -i/2$ and $G_{11}^A(t_1,t_1)= i/2$ considering $\theta(t=0)=1/2$ and in the short time limit one can approximate $G_{11}^K(t_1,t_1)\approx g_{11}^{K} = 2i \langle c_1^{\dagger} c_1 \rangle_{t_1=0} - i$ which for initially vaccum state reduces to $G_{11}^K(t_1,t_1) \approx -i$. As a result, the short-time limit generating function reduces to
\begin{equation}
\chi(\lambda, t) = \Gamma \, t \, \big(e^{i \lambda}-1\big) + O(t^2)
\end{equation}
which once again matches with Eq.~\eqref{short} in the main text.

\section{Expression for $\langle N(t) \rangle$ for arbitrary time}
    
In this section, we derive the analytical expression for the first  cumulant i.e., $\langle N(t)\rangle$ starting from Eq.~\eqref{CGF-general-series}. $\langle N(t)\rangle$ can be obtained from the CGF $\chi(\lambda)$ by taking first derivatives $\kappa_1(t)$ with respect to $\lambda$ and the setting $\lambda=0$. 
\begin{equation}
\kappa_1(t)=
\left.
\frac{\partial \chi(\lambda,t)}
{\partial (i\lambda)}
\right|_{\lambda=0},
\label{eq:kappa-sp}
\end{equation}
It is easy to see that in the $\log$ series in Eq.~\eqref{CGF-general-series}, only the first term  contributes to first culumant $\kappa_1$, 
The average number of fermions $\langle N(t) \rangle$ at arbitrary time instant $t$ is given by
\begin{eqnarray}
\langle N(t) \rangle &=&
 \mathrm{Tr}_{j,t} \Big[{\breve{G}_0} \, \frac{\partial}{\partial (i \lambda)}\breve{{\Sigma}}_{\lambda}\Big{|}_{\lambda=0}\Big] \nonumber \\
 && =\int_{t_0}^{t} dt_1 \, \int_{t_0}^{t} dt_2 \, \mathrm{Tr}_{j}\Big[\breve{{G}}(t_1,t_2)\,\frac{\partial}{\partial (i \lambda)}\breve{{\Sigma}}_{\lambda}(t_2,t_1)\Big{|}_{\lambda=0} \Big].
 \end{eqnarray}
Performing matrix multiplication and further simplification leads to 
\begin{equation}
\langle N(t) \rangle = \frac{i \Gamma}
{2} \int_{t_0}^{t} dt_1 \Big[G_{11}^R(t_1,t_1)-G_{11}^A(t_1,t_1) + G_{11}^K(t_1,t_1)\Big].
\end{equation}
Following the definitions of the Green's functions in Eq.~\eqref{Gr-corr}, Eq.~\eqref{Ga-corr}, and Eq.~\eqref{Gk-corr}, we note that $G_{11}^R(t_1,t_1)= -i/2$ and $G_{11}^A(t_1,t_1)= i/2$. Also for the Keldysh component,  $i G_{11}^K(t_1,t_1)= 1 - 2 \, \langle c_1^{\dagger}(t_1) c_1 (t_1) \rangle$, we obtain
\begin{equation}
\langle N(t) \rangle  = {\Gamma} \int_{t_0}^{t} dt_1 \Big[1 + i \, G_{11}^{<}(t_1,t_1)\Big] = {\Gamma} \int_{t_0}^{t} dt_1 \Big[1- \langle n_1(t_1)\rangle\Big]
\label{Nt-supp}
\end{equation}
where $G_{11}^{<}(t,t)= i \langle c_1^{\dagger}(t) c_1(t) \rangle = i \langle n_1(t)\rangle $ with $\langle n_1(t)\rangle$ being the local density at the injected site $1$. For the non-interacting lattice setup, the Dyson equation for the lesser component can be expressed as 
\begin{equation}
\label{Gless-diff}
G^{<}(\tau,\tau')= g^<(\tau,\tau')+ i \Gamma \int d\tau_1 \, G^R(\tau,\tau_1)  \, P_1 \, G^A(\tau_1,\tau'),
\end{equation}
which implies for the first component,
\begin{equation}
\label{Gless}
G^{<}_{11}(t,t')= g_{11}^{<}(t,t')+ i \Gamma \int_{t_0}^{t} dt_1 \, G_{11}^R(t,t_1)  \, G_{11}^A(t_1,t').
\end{equation}
To evaluate $\langle N(t)\rangle$, we want the lesser correlator $G^{<}_{11}(t,t')$ at the same time $t=t'$. For the empty initial state $g_{11}^{<}(t,t)=0$ and hence 
\begin{equation}
\label{Gless-reduced}
G^{<}_{11}(t,t)= + i \Gamma \int_{t_0}^{t} dt_1 \, G_{11}^R(t,t_1)  \, G_{11}^A(t_1,t).
\end{equation}
Following Eq.~\eqref{Nt-supp}, we can then write 
\begin{equation}
\langle N(t) \rangle  = {\Gamma} \int_{t_0}^{t} dt_1 \Bigg[1 - \Gamma \int_{t_0}^{t_1} dt_2 \Big[G_{11}^R(t_1,t_2)  \, G_{11}^A(t_2,t_1)\Big]\Bigg].
\end{equation}
Now following Eq.~\eqref{Gr-diff} and Eq.~\eqref{Ga-diff}, which are differential equations that retarded and advanced Green's function satisfy, one can write down the solutions as
\begin{eqnarray}
G^R(t,t') &=& -i \, \theta(t-t') \, e^{-ih_{\rm e}(t-t')} \\
G^A(t,t') &=& +i \, \theta(t'-t) \, e^{-ih^{\dagger}_{\rm e}(t-t')}
\end{eqnarray}
where $h_{\rm e}= h - i \frac{\Gamma}{2} P_1$. We also note that  $G^A(t,t')= \big[G^R(t',t)]^{\dagger}$. Since these Green's functions are time-translationally invariant, we can finally write
\begin{equation}
\langle N(t) \rangle  = {\Gamma} \int_{t_0}^{t} dt_1 \Bigg[1 - \Gamma \int_{t_0}^{t_1} dt_2 |G_{11}^R(t_2)|^2 \Bigg].
\end{equation}
where 
\begin{eqnarray}
|G_{11}^R(t)|^2 &=&  |\langle 1| e^{-ih_{\rm e} t}|1\rangle|^2.
\end{eqnarray}
This expression resembles Eqs.~(34) and (35) derived in Ref.~\cite{Krapivsky_2019}.

\section{Local retarded Green's function and lattice self-energy from non-interacting lattices}

We next obtain the local retarded Green's function at the injection
site $G_{11}^R[\omega]$ by separating the first site from the remaining lattice. The
single-particle Hamiltonian $h_{ij}$ can be written in the block form
\begin{equation}
h=
\begin{pmatrix}
\epsilon_1 & \mathbf{v}\\
\mathbf{v}^T & h_{\rm lat}
\end{pmatrix},
\label{eq:block-hamiltonian}
\end{equation}
where $\epsilon_1$ is the onsite energy of the injection site,
$h_{\rm lat}$ denotes the single-particle Hamiltonian of the
remaining sites $2,\ldots,N$, and $\mathbf{v}$ contains the hopping
amplitudes connecting site $1$ to the rest of the lattice. In the presence of particle injection at site $1$, the retarded Green's function is
\begin{equation}
G^R[\omega]
=
\left[
\omega I-h+i\frac{\Gamma}{2}P_1
\right]^{-1},
\qquad
P_1=|1\rangle\langle1|.
\label{eq:full-retarded-GF}
\end{equation}
Using the block decomposition in Eq.~\eqref{eq:block-hamiltonian},
    the inverse retarded Green's function takes the form
\begin{equation}
\big[G^R[\omega]\big]^{-1}
=
\begin{pmatrix}
\omega-\epsilon_1+i\Gamma/2
&
-\mathbf{v} \\
-\mathbf{v}^T
&
\omega I_{\rm lat}-h_{\rm lat}+i0^+
\end{pmatrix}.
\label{eq:block-inverse-GR}
\end{equation}
For a generic block matrix
\begin{equation}
M=
\begin{pmatrix}
A&B\\
C&D
\end{pmatrix},
\end{equation}
the upper-left block of its inverse is given by the inverse of the
Schur complement of $D$,
\begin{equation}
[M^{-1}]_{11}
=
\left(A-BD^{-1}C\right)^{-1}.
\label{eq:schur-complement}
\end{equation}
Applying Eq.~\eqref{eq:schur-complement} to
Eq.~\eqref{eq:block-inverse-GR}, we obtain
\begin{equation}
G_{11}^R[\omega]=
\Big[
\omega-\epsilon_1+i\Gamma/2-\mathbf{v} 
\left(
\omega I_{\rm lat}-h_{\rm lat}+i0^+
\right)^{-1}
\mathbf{v}^T
\Big]^{-1}.
\label{eq:G11-schur}
\end{equation}
Introducing the retarded Green's function of the lattice with
site $1$ removed,
\begin{equation}
g_{\rm lat}^R[\omega]
=
\left[
\omega I_{\rm lat}-h_{\rm lat}+i0^+
\right]^{-1},
\label{eq:self-GF}
\end{equation}
we define the corresponding lattice self-energy as
\begin{equation}
\Sigma_{\rm lat}^R[\omega]
=
\mathbf{v} \,
g_{\rm lat}^R[\omega]\,
\mathbf{v}^T.
\label{eq:lattice-self-energy}
\end{equation}
The local retarded Green's function $G_{11}^R[\omega]$ at the injection site therefore
takes the compact form
\begin{equation}
G_{11}^R[\omega]
=
\frac{1}{
\omega+i\Gamma/2
-\Sigma_{\rm lat}^R[\omega]
}.
\label{eq:G11-final}
\end{equation}
The corresponding advanced quantities are obtained as
\begin{equation}
\Sigma_{\rm lat}^A[\omega]
=
\left[\Sigma_{\rm lat}^R[\omega]\right]^*,
\qquad
G_{11}^A[\omega]
=
\left[G_{11}^R[\omega]\right]^*,
\end{equation}
for a Hermitian lattice Hamiltonian.

\subsection{Lattice self-energy for a nearest-neighbor tight-binding chain}

For the nearest-neighbor tight-binding model, the lattice self-energy given in Eq.~\eqref{eq:self-GF} can be calculated analytically. We consider the Hamiltonian
\begin{equation}
H
=
J\sum_{j=1}^{\infty}
\left(
c_j^\dagger c_{j+1}
+
c_{j+1}^\dagger c_j
\right),
\label{eq:TB-Hamiltonian}
\end{equation}
where we set the onsite energies to be  zero. Upon separating
the injection site, $j=1$, from the remaining semi-infinite
lattice, the vector $\mathbf{v}$ takes a simple form
\begin{equation}
\mathbf{v}
=
\begin{pmatrix}
J & 0 & 0 & \cdots
\end{pmatrix}.
\end{equation}
As a result, the retarded lattice self-energy simplifies to
\begin{equation}
\Sigma_{\rm lat}^R[\omega]
=
J^2 g_s^R[\omega],
\label{eq:Sigma-TB}
\end{equation}
where $g_s^R[\omega]$ denotes the surface retarded Green's
function of the remaining semi-infinite tight-binding chain. The surface Green's function can be obtained recursively. After
removing the first site of the semi-infinite chain, the remaining
system is identical to the original semi-infinite chain. Therefore,
\begin{equation}
g_s^R[\omega]
=
\frac{1}{
\omega+i0^+-J^2g_s^R[\omega]
}.
\label{eq:surface-recursion}
\end{equation}
Equation~\eqref{eq:surface-recursion} gives a quadratic equation
\begin{equation}
J^2\left[g_s^R[\omega]\right]^2
-
(\omega+i0^+)g_s^R[\omega]
+1
=
0.
\end{equation}
Its solution is
\begin{equation}
g_s^R[\omega]
=
\frac{
\omega+i0^+
\pm
\sqrt{(\omega+i0^+)^2-4J^2}
}{
2J^2
}.
\end{equation}
The appropriate solution for the retarded Green's function can be decided by the conditions that  $g_s^R[\omega]$ decays for large $\omega$ and
${\rm Im}\,g_s^R[\omega]\leq0$. Thus,
\begin{equation}
g_s^R[\omega]
=
\frac{
\omega-
\sqrt{(\omega+i0^+)^2-4J^2}
}{
2J^2
}.
\label{eq:surface-GF-TB}
\end{equation}
Using Eq.~\eqref{eq:Sigma-TB}, the corresponding lattice
self-energy is
\begin{equation}
\Sigma_{\rm lat}^R[\omega]
=
\frac{1}{2}
\left[
\omega-
\sqrt{(\omega+i0^+)^2-4J^2}
\right].
\label{eq:Sigma-TB-final}
\end{equation}
Inside the tight-binding band, $|\omega|<2J$, this reduces to
\begin{equation}
\Sigma_{\rm lat}^R[\omega]
=
\frac{\omega}{2}
-
\frac{i}{2}
\sqrt{4J^2-\omega^2},
\qquad |\omega|<2J.
\label{eq:Sigma-TB-inside}
\end{equation}
The corresponding spectral broadening $\Gamma_{\rm lat}[\omega]$ induced  by the remaining lattice is therefore
\begin{align}
\Gamma_{\rm lat}[\omega]
=
i\left[
\Sigma_{\rm lat}^R[\omega]
-
\Sigma_{\rm lat}^A[\omega]
\right]
=
-2\,{\rm Im}\,\Sigma_{\rm lat}^R[\omega]
=
\sqrt{4J^2-\omega^2},
\qquad |\omega|<2J.
\label{eq:Gamma-TB}
\end{align}
Outside the tight-binding band,
$\Gamma_{\rm lat}[\omega]=0$.
Substituting Eq.~\eqref{eq:Sigma-TB-inside} into the expression
for the local retarded Green's function at the injection site,
\begin{equation}
G_{11}^R[\omega]
=
\frac{1}{
\omega+i\Gamma/2-\Sigma_{\rm lat}^R[\omega]
},
\end{equation}
we obtain, for $|\omega|<2J$,
\begin{equation}
G_{11}^R[\omega]
=
\frac{2}{
\omega+
i\left[
\Gamma+\sqrt{4J^2-\omega^2}
\right]
}.
\label{eq:G11-TB}
\end{equation}
Consequently, the effective transmission function entering the
long-time CGF,
\begin{equation}
{\cal T}[\omega]
=
\Gamma\,
\Gamma_{\rm lat}[\omega]
\left|G_{11}^R[\omega]\right|^2,
\end{equation}
takes the explicit form
\begin{equation}
{\cal T}[\omega]
=
\frac{
4\Gamma\sqrt{4J^2-\omega^2}
}{
\omega^2+
\left[
\Gamma+\sqrt{4J^2-\omega^2}
\right]^2
},
\qquad |\omega|<2J,
\label{eq:T-TB}
\end{equation}
and ${\cal T}[\omega]=0$ outside the band. The long-time CGF
therefore becomes
\begin{equation}
\chi(\lambda,t)
=
T
\int_{-2J}^{2J}
\frac{d\omega}{2\pi}
\ln
\left[
1+
{\cal T}[\omega]
\left(e^{i\lambda}-1\right)
\right].
\label{eq:CGF-TB-explicit}
\end{equation}

\subsection{Lattice self-energy for long-range lattice model}
For a long-range lattice model with power-law hopping, we write the single particle Hamiltonian matrix elements as 
\begin{equation}
h_{ij}
=
\frac{J}{|i-j|^\alpha},
\qquad i\neq j
\end{equation}
and once again the onsite energies are set to zero. The vector $\mathbf{v}$ between the injection site and the remaining
lattice is
\begin{equation}
\mathbf{v}
=
J
\begin{pmatrix}
1 &
2^{-\alpha} &
3^{-\alpha} &
\cdots
\end{pmatrix}.
\label{eq:long-range-v}
\end{equation}
Consequently, the lattice self-energy can be written explicitly as
\begin{equation}
\Sigma_{\rm lat}^R[\omega]
=
J^2
\sum_{m,n=1}^{L-1}
\frac{
[g_{\rm lat}^R [\omega]]_{mn}
}{
m^\alpha n^\alpha
}
\label{eq:long-range-self-energy}
\end{equation}
In the thermodynamic limit, $N\rightarrow\infty$, the remaining
lattice becomes semi-infinite and Eq.~\eqref{eq:long-range-self-energy}
defines the corresponding self-energy seen by the injection site.

\begin{figure}
\includegraphics[width=0.8\textwidth]{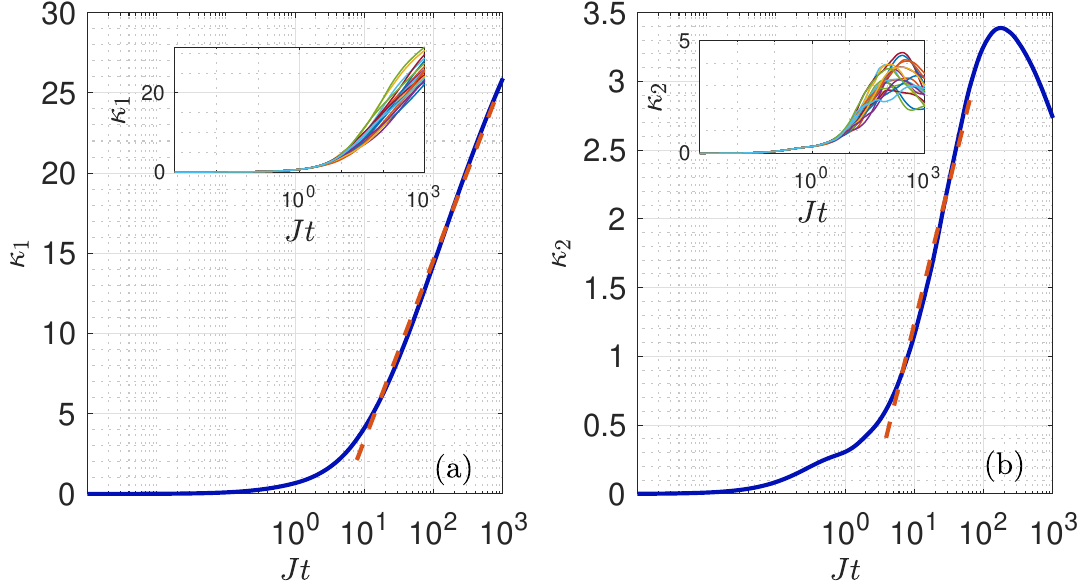}
\caption{Plots for quantum dynamics of particle number growth for first two cumulants: (a) $\kappa_1$, (b) $\kappa_2$ for a disordered non-interacting lattice, described by the Hamiltonian in Eq.~\eqref{eq:ham_disorder}. Due to the existence of localized states, the long-time growth is not linear in time, in contrast to what is observed for clean non-interacting lattice models. In fact the first cumulant $\kappa_1$ shows $\ln t$ growth over three decades and second cumulant $\kappa_2$ shows $\ln t$ growth over two decades . The insets in (a) and (b) shows results for different disorder realizations. The parameters used in the simulation are $L=50$, $\Gamma=1$, $W=2$, and the disorder average is done over 20 different realizations.}
\label{fig:disorder}
\end{figure}

\section{Results for first and second cumulants of particle number growth for one-dimensional disorder lattice}

In this section, we provide results for a case where the long-time growth of the cumulants for injected particles is \textit{not} linear in time. Recall that Eq.~\eqref{eq:CGF-long-time-general}, which provides linear in time growth of cumulants in the asymptotic regime is strictly valid when the underlying non-interacting lattice possess delocalized states. Such a growth is typically not expected when the lattice hosts localized or bound states which have finite overlap with the injection site. In that case,  memory of the initial condition and oscillatory contributions to the local Green's function sustains and the system can exhibit growth that is very different from linear. 

In order to show this explicitly, we consider a disordered non-interacting lattice that hosts localized states. The Hamiltonian is given by 
\begin{equation}
H= \sum_{j=1}^{L-1} \Big[
J \left( c_j^{\dagger} c_{j+1}
+ c_{j+1}^{\dagger} c_j \right) + \epsilon_j c_j^{\dagger} c_j \Big]
,
\label{eq:ham_disorder}
\end{equation}
where $J>0$ is the hopping strength and $\epsilon_i$ is the onsite disorder and 
chosen uniformly from a random distribution $[-W/2,W/2]$, where $W$ characterizes the strength of the disorder.

In Fig.~\ref{fig:disorder}, we plot the first two cumulants $\kappa_1$, $\kappa_2$ and as expected, due to the existence of localized states, the long-time growth of the cumulants is not linear in time. In fact the first cumulant $\kappa_1$ shows $\ln t$ growth over three decades in time, whereas, the second cumulant also shows a good $\ln t$ growth (over two decades) before it reaches maximum and drops subsequently.


\end{document}